\documentclass[]{aa}
\usepackage[varg]{txfonts}
\usepackage{graphicx}
\usepackage{subfig}
\usepackage{stackengine}
\usepackage[pscoord]{eso-pic}
\usepackage[export]{adjustbox}
\bibpunct{(}{)}{;}{a}{}{,} 

\title{Inflows towards active regions and the modulation of the solar cycle: a parameter study}
\author{D. Martin-Belda\inst{\ref{inst1}} \inst{\ref{inst2}}
\and R. H. Cameron\inst{\ref{inst1}}
}

\institute{Max-Planck-Institut für Sonnensystemforschung, Justus-von-Liebig-Weg 3, 37077 Göttingen, Germany\label{inst1} \and
Institut für Astrophysik, Georg-August-Universität Göttingen, 37077 Göttingen, Germany\label{inst2}
}

\date{Received / Accepted}


\abstract
{}
{We aim to investigate how converging flows towards active regions affect the surface transport of magnetic flux, as well as their impact on the generation of the Sun's poloidal field. The inflows constitute a potential non-linear mechanism for the saturation of the global dynamo and may contribute to the modulation of the solar cycle in the Babcock-Leighton framework.}
{We build a surface flux transport code incorporating a parametrized model of the inflows and run simulations spanning several cycles. We carry out a parameter study to assess how the strength and extension of the inflows affect the build-up of the global dipole field. We also perform simulations with different levels of activity to investigate the potential role of the inflows in the saturation of the global dynamo.}
{We find that the interaction of neighbouring active regions can lead to the occasional formation of single-polarity magnetic flux clumps inconsistent with observations. We propose the darkening caused by pores in areas of high magnetic field strength as a plausible mechanism preventing this flux-clumping. We find that inflows decrease the amplitude of the axial dipole moment by a $\sim30\,\%$, relative to a no-inflows scenario. Stronger (weaker) inflows lead to larger (smaller) reductions of the axial dipole moment. The relative amplitude of the generated axial dipole is about $9\%$ larger after very weak cycles than after very strong cycles. This supports the inflows as a non-linear mechanism capable of saturating the global dynamo and contributing to the modulation of the solar cycle within the Babcock-Leighton framework.}
{}

\begin{document}
\titlerunning{Inflows and random velocities}
\authorrunning{D. Martin-Belda \& R. H. Cameron}
\maketitle

\section{Introduction}\label{sec:intro}
The magnetic activity of the Sun follows an 11-year cycle. At the time of minimum activity, the surface magnetic field is concentrated at the polar caps and presents a strongly dipolar configuration. As the cycle progresses, new magnetic flux erupts in the form of bipolar magnetic regions (BMRs). The preceding polarity (relative to the Sun's sense of rotation) of the new BMRs tends to emerge closer to the equator (Joy's law), and is of the same sign as the polar field in the same hemisphere at the immediately previous activity minimum (Hale's law). The latitudinal separation of polarities favors the cross-equatorial transport of preceding polarity flux, which causes the gradual cancellation and eventual reversal of the polar fields. When the next activity minimum is reached, the global field is again nearly dipolar and reversed with respect to the previous activity minimum. The full magnetic cycle is therefore 22 years long. These activity cycles show pronounced variability, both cycle to cycle and on longer time scales \citep[for a review of the solar cycle, see][]{hathaway2015solarcycle}.

It has been shown that the strength of the polar fields at activity minima strongly correlates with the amplitude of the subsequent cycle \citep[see, {e.g.}][]{schatten1978using, choudhuri2008prospects, wang2009precursors,  munoz2013solar}. This is supportive of the Babcock-Leighton model of solar dynamo, in which the polar fields at activity minima represent the poloidal field threading the Sun, from which the toroidal field of the next cycle is generated, rather than being a secondary manifestation of a dynamo mechanism operating below the surface \citep{cameron2015crucial}. It follows that, in this framework, an activity-related feedback mechanism affecting the surface transport of magnetic flux could provide a means for saturating the dynamo by limiting the build-up of the polar fields (and therefore the regeneration of the poloidal field) and possibly contribute also to the observed variability of the cycle amplitude. One candidate for such a mechanism are the near-surface, converging flows towards active regions \citep{cameron2012strengths}. These flows, first reported by \cite{gizon2001probing}, have magnitudes of $\sim 50\,\mathrm{m\,s^{-1}}$ and can extend up to $30\,\mathrm{^\circ}$ away from the center of the active region. The inflows are possibly driven by the temperature gradient arising from the enhanced radiative loss in areas of strongly concentrated magnetic field \citep{spruit2003torsional,gizon2008observation}.

The question of how these inflows affect the surface transport of magnetic flux and the build-up of the polar fields has been addressed in a number of works. Their main effect is the limitation of the latitudinal separation of the polarities of BMRs, which causes a reduction of the global dipole with respect to a no-inflows scenario \citep{jiang2010effect}. This effect dominates in strong cycles, while in weaker cycles the inflows driven by low-latitude BMRs mainly enhance the cross-equatorial transport of magnetic flux, resulting in stronger polar fields. The inclusion of the inflows in surface flux transport simulations improves the correlations of the amplitude of the global dipole with the inferred open heliospheric flux in cycles 13 to 21 \citep{cameron2012strengths}, but produces a weaker match with the observed butterfly diagram and dipole reversal times in cycles 23 and 24 \citep{yeates2014coronal}.

All the studies cited above modeled the inflows as an axisymmetric perturbation of the meridional flow converging toward the activity belts. \cite{derosa2006consequences} included a more realistic model of inflows in their surface flux transport model, but the converging flows severely affected the dispersal of magnetic flux in their simulations, leading to unrealistic clumping of magnetic flux in spite of diffusion by supergranules. In a recent work \citep{mbelda2015inflows}, we studied the impact of the inflows in the evolution of an isolated BMR, and showed that turbulent diffusion and differential rotation are sufficiently strong to counteract the converging flows, which decline quickly due to flux cancellation. A probable reason for the discrepancy with the aforementioned study is the additional damping of the turbulent diffusivity inside active regions that these authors included to match observations. We argued that the inflows alone can cause this effect.

In this work we go on studying the effect of the inflows on the surface transport of magnetic flux. Our main question is the impact realistic, non axisymmetric inflows may have on the generation of the large-scale poloidal field. As mentioned above, this could provide a non-linear saturation mechanism for the global dynamo and contribute to the solar cycle variability. A second problem concerns whether our previous result on the dispersal of flux against converging flows holds in global simulations. To address these questions, we incorporated two non-axisymmetric parametrizations of the inflows in a surface flux transport model. The paper is structured as follows: we first introduce our model (Sec. \ref{sec:SFTmodel}); then, we examine a case with inflows whose strength and extension are compatible with observations (Sec. \ref{sec:refCase}); next, we carry out a parameter study to test how these two magnitudes, as well as the activity level, may affect the build-up of the global magnetic dipole at activity minima (Sec. \ref{sec:paramStudy}); the results are summarized and briefly discussed (Sec. \ref{sec:disConc}).

\section{Surface Flux Transport Model}\label{sec:SFTmodel}
\subsection{Surface flux transport equation}\label{sec:SFTE}
The evolution of the magnetic field on the solar surface is governed by the radial component of the induction equation \citep{devore1984meridional}:
\begin{align}
	\frac{\partial B_r}{\partial t} &= - \frac{1}{R_\odot \sin \theta}\left[\frac{\partial (B_r u_\theta \sin \theta)}{\partial \theta}+\frac{\partial (B_r u_\phi)}{\partial \phi}\right]\nonumber\\
	& + \frac{\eta}{R_\odot^2}\left[\frac{1}{\sin\theta}\frac{\partial}{\partial\theta}\left(\sin\theta\frac{\partial B_r}{\partial \theta}\right)+\frac{1}{\sin^2\theta}\frac{\partial^2 B_r}{\partial \phi^2}\right]\label{eq:sft}\\
	& + S(\theta,\phi,t).\nonumber
\end{align}
where $\phi$ and $\theta$ denote solar longitude and colatitude, respectively. The first term on the right hand side represents the advection of magnetic flux by the surface flows, which include differential rotation, meridional flow, and inflows towards active regions:
\begin{align}
u_\theta &= v_m(\theta) + w_\theta(\theta,\phi);\\
u_\phi &= R_\odot \sin\theta \, \Omega(\theta) + w_\phi(\theta,\phi).
\end{align}
Here, $v_m$ is the velocity of the meridional flow, $\Omega(\theta)$ is the angular velocity of the differential rotation and $w_\phi$ and $w_\theta$ are the components in spherical coordinates of the inflows.

We adopt the differential rotation profile from \cite{snodgrass1983magnetic}:
\begin{equation}
\Omega(\theta) = 13.38 - 2.30\cos^2\theta - 1.62\cos^4\theta\;\mathrm{[^\circ/day]}.
\end{equation}

Following \cite{vanBallegooijen1998fchannels}, we model the meridional flow as:
\begin{equation}
v_m(\lambda) = \begin{cases}
11\sin(2.4\lambda)\;\mathrm{[m/s]} &\text{if $|\lambda|< \lambda_0$};\\
0 &\text{if $|\lambda|\geq \lambda_0$},
\end{cases}
\label{eq:mFlow}
\end{equation}
where $\lambda$ denotes solar latitude and $\lambda_0=75^\circ$.

\begin{figure}[t]
  \centering
  \resizebox{\hsize}{!}{\includegraphics[width=\textwidth]{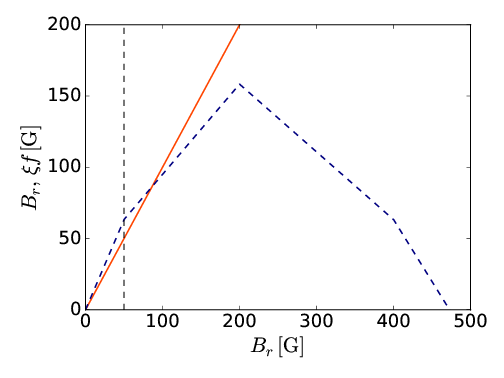}}
  \caption{$B_r$ (orange line) and $\xi f$ (dashed, blue line) as a function of $B_r$. The dashed vertical line marks $B_r=50\,\mathrm{G}$.}
  \label{fig:calib}
\end{figure}

The second term on the right hand side of Eq. \eqref{eq:sft} describes the flux dispersal by convective flows as a random walk/diffusion process \citep{leighton1964transport}. We choose $\eta=250\,\mathrm{km^2/s}$, a value in agreement with observations \citep{schrijver1990patterns,jafarzadeh2014migration} 
and consistent with the evolution of the large scale fields \citep{cameron2010surface}.

The term $S(\theta,\phi,t)$ describes the emergence of new active regions, and is described in detail in \cite{baumann2004parameter}. The synthetic activity cycles in our simulations are $13$ years long, with a two-years overlap between cycles, so the time distance between consecutive cycle minima is $11\,\mathrm{years}$. The activity level (the number of new BMRs per day) is governed by a Gaussian function whose height peaks halfway into the the cycle. At the beginning of the cycle, the BMRs emerge at a mean latitude of $40^\circ$ with a standard deviation of $10^\circ$. These values decrease linearly and reach a mean latitude of $5^\circ$ and a standard deviation of $5^\circ$ at the end of the cycle. We do not consider active longitudes in this study, so the random distribution is uniform in $\phi$.
Following \cite{vanBallegooijen1998fchannels}, we represent a BMR by two circular patches of opposite polarity. The magnetic field of each patch is given by:
\begin{equation}
B_r(\theta,\phi) = B_{max}\left(\frac{\delta}{\delta_0}\right)^2 \exp\left\{ -\frac{2[1-\cos\beta_{\pm}(\theta,\phi)]}{\delta_0^2}\right\},
\label{eq:newBmrEarlyDiff}
\end{equation}
where $\beta_{\pm}$ is the heliocentric angle between the center of the $(\pm)$ polarity patch and the surface point $(\theta,\phi)$; $\delta$ denotes the angular size of the $BMR$ and $\delta_0 = 4^\circ$. The size of the BMRs follows a distribution $n(\delta)\propto\delta^4$. This distribution was derived by \cite{schrijver1994budget} from observations for BMRs with sizes ranging from $3.5^\circ$ to $10^\circ$. BMRs smaller than $3.5^\circ$ cannot be well resolved in our simulations, so they are assumed to diffuse without interacting with the rest of the flux until they reach this size. The maximum field strength upon emergence, $B_{max}$, is adjusted so the total flux input per cycle is $\sim 8.9\cdot10^{24}\,\mathrm{Mx}$ \citep{schrijver1994budget}.

\subsection{Parametrization of the inflows}\label{sec:inflowsPar}

\begin{figure*}[t]
  \centering
  \resizebox{\hsize}{!}{\includegraphics{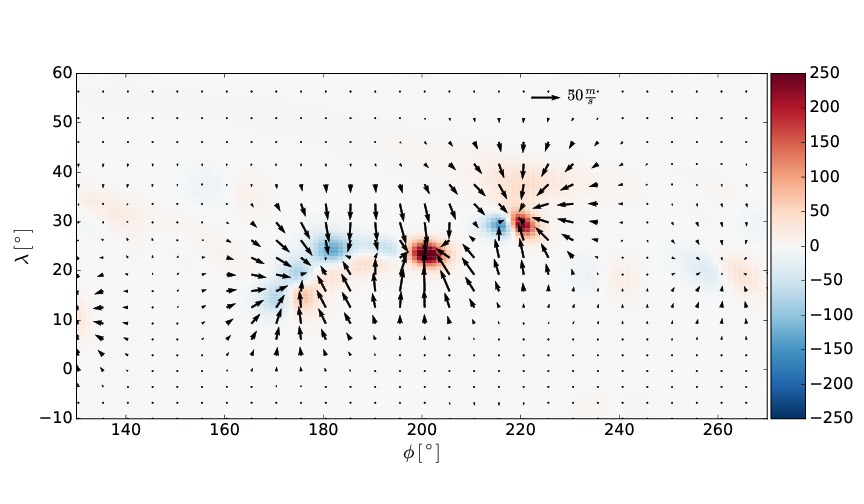}}
  \caption{Model of inflows towards an activity complex formed by several emerged BMRs. The color scale encodes the strength of the magnetic field $B_r$ in Gauss, and saturates at $250\,\mathrm{G}$. Red and blue indicate opposite polarities. The values of the parameters in Eq. \eqref{eq:infVelocParamB} are $a=1.8\cdot10^8\,\mathrm{m^2 G^{-1} s^{-1}}$ and $\mathrm{FWHM = 15^\circ}$.}
  \label{fig:inflowsExample}
\end{figure*} 

We test two different models of inflows. The first one is based upon the parametrization of \cite{cameron2012strengths},
\begin{equation}
\mathbf{w}=a \, \nabla \left(\frac{\cos\lambda}{\cos 30^\circ}|\hat{B_r}|\right),
\label{eq:infVelocParamB}
\end{equation}
where $|\hat{B_r}|$ is the absolute value of the magnetic field smoothed with a Gaussian. Adjusting the full width at half maximum (FWHM) of the Gaussian allows us to control the extension of the inflows. Note that the gradient of the smoothed magnetic field generally decreases with increasing width of the Gaussian. Hence, for a fixed value of $a$, wider inflows are weaker. The factor $\cos\lambda/\cos 30^\circ$ is introduced to quench unrealistically strong poleward flows arising from the gradient of the polar fields. Figure \ref{fig:inflowsExample} shows the inflows around an activity complex for our reference values $a=1.8\cdot10^8\,\mathrm{m^2/G\,s}$ and $\mathrm{FWHM}=15^\circ$. The value of $a$ is chosen so that the peak inflow velocity around an isolated BMR of size $10^\circ$ is $\sim50\,\mathrm{m/s}$, in agreement with observations. We will hereafter refer to this parametrization as the $B-$parametrization.

The second parametrization of the inflows is motivated by the results of \cite{voegler2005brightness}. The author's radiative MHD simulations suggest that the relation between the average magnetic field in an active region and the integrated radiation flux is non-monotonic, peaking at about $\sim 200\,\mathrm{G}$. For stronger average fields, the formation of dark pores reduces the radiation output. This can effectively reduce the radiative cooling in active regions, and thus limit the strength of the inflows. We attempt to capture this effect by substituting $B_r$ in Eq. \eqref{eq:infVelocParamB} with the angle integrated radiation flux normalized to the quiet-sun value (which we denote by $f$), taken from the left panel of Fig. 2 in \cite{voegler2005brightness},
\begin{equation}
\mathbf{w}=\xi a \, \nabla \left(\frac{\cos\lambda}{\cos 30^\circ}\hat{f}\right).
\label{eq:infVelocParamF}
\end{equation}
The prefactor $\xi=6.3\cdot10^4$ was adjusted such that the peak inflow velocity around an isolated, $10^\circ-$sized BMR is $\sim 50\,\mathrm{m/s}$ for our reference values of $a$ and the FWHM. This parametrization is referred to as $f-$parametrization in the remainder of the paper. The $f-$parametrization produces weaker inflow velocities in regions of strong magnetic field. However, since the slope of $f$ is steeper than the slope of $B$ between $0$ and $50\,\mathrm{G}$ (see Fig. \ref{fig:calib}), the contribution to the inflows of areas with fields lower than $50\,\mathrm{G}$ value to the inflows will be stronger than in the $B-$case.

One of the problems we address in this paper is the suppression of magnetic flux dispersal in the presence of inflows found by \cite{derosa2006consequences}. These authors parametrize the inflows in the following way:

\begin{equation}
    \mathbf{w} = a\nabla |B_r|^b
    \label{eq:deRosaParam}
\end{equation}

This parametrization (with $b=1$) is the same as ours except for the geometric factor similar to the one introduced in \cite{cameron2012strengths} to prevent strong inflows near the poles.

\begin{figure*}[t]
  \centering
  \begin{adjustbox}{minipage=\textwidth,scale=0.94}
  \subfloat{
    \subfloat{
      \stackunder{\includegraphics[width=.323\textwidth]{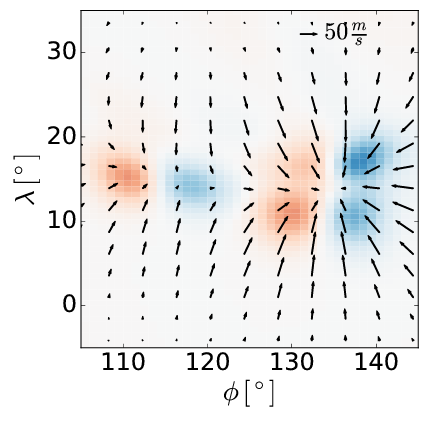}}{\hspace{0.8cm}$t=0$}
    }
    \subfloat{
      \stackunder{\includegraphics[width=.316\textwidth]{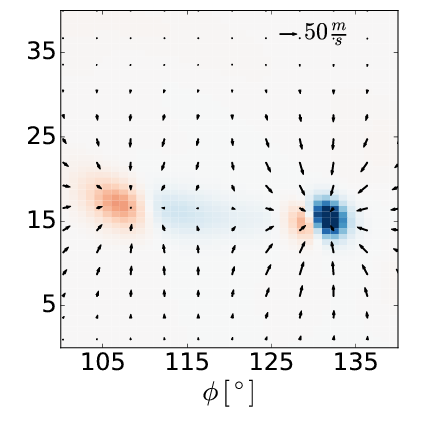}}{\hspace{0.2cm}$t=27\,\mathrm{days}$}
    }
    \subfloat{
      \stackunder{\includegraphics[width=.361\textwidth]{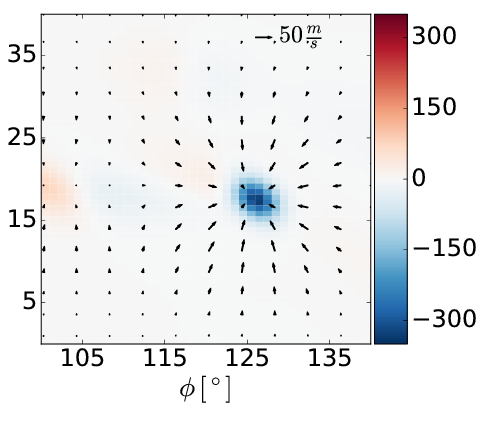}}{\hspace{-1cm}$t=54\,\mathrm{days}$}
    }
  }\\
  \subfloat{
    \subfloat{
      \stackunder{\includegraphics[width=.323\textwidth]{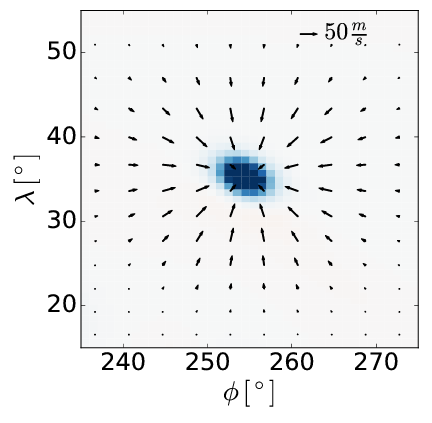}}{\hspace{0.85cm}$t=1.2\,\mathrm{years}$}
    }
    \subfloat{
      \stackunder{\includegraphics[width=.316\textwidth]{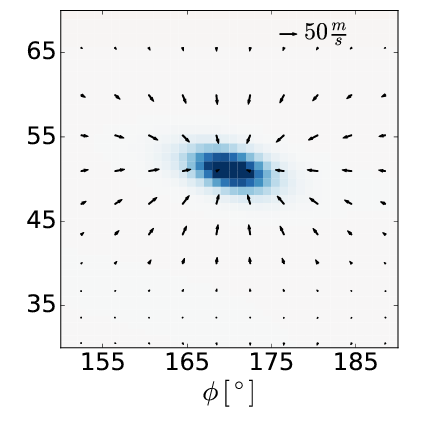}}{\hspace{0.37cm}$t=2.0\,\mathrm{years}$}
    }
    \subfloat{
      \stackunder{\includegraphics[width=.361\textwidth]{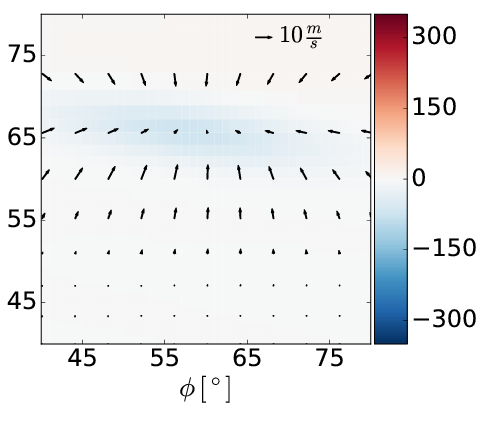}}{\hspace{-1.cm}$t=3.0\,\mathrm{years}$}
    }
  }
  \end{adjustbox}
  \caption{Time series showing the evolution of a long-standing, single-polarity magnetic flux clump. Time progresses from left to right and from top to bottom. The color scale indicates the magnetic field strength in Gauss, and saturates at $350\,\mathrm{G}$. The arrows represent the strength and direction of the inflows.}
  \label{fig:tSeries}
  
\end{figure*}

\subsection{Numerical treatment}\label{sec:numTreatment}

In order to integrate Eq. \eqref{eq:sft} we developed a surface flux transport code. The equation is expressed in terms of $x=\cos\theta$. We calculate the $x-$derivative in the advection term with a fourth-order centered finite differences scheme. The derivative in the $\phi$ direction is calculated in Fourier space. We use a fourth-order Runge-Kutta scheme to advance the advection terms in time. 
The diffusion term in the $x-$direction is treated with a Crank-Nicolson scheme. We treat the $\phi-$diffusion term by computing the exact solution of the diffusion equation for the Fourier components of $B_r$.

We validated the code by reproducing the results for the reference case of the study by \cite{baumann2004parameter}.

Calculating the inflows requires smoothing the absolute value of the magnetic field (or the normalized radiation flux, in the $f-$ parametrization) with a Gaussian of given $FWHM$ every time step. The result of this operation is identical to diffusing $|B_r|$ (or $f$) with a given diffusion coefficient $\eta_s$ for a time interval $\Delta t'$, related to the $FWHM$ of the smoothing Gaussian through:
\begin{equation}
FWHM = 4\sqrt{\Delta t'\,\eta_s\,\ln2}.
\label{eq:smoothDiff}
\end{equation}
We emphasize that the time $t'$ in Eq. \eqref{eq:smoothDiff} is different from the simulation time $t$. The diffusion of $|B_r|$ or $f$ is a mathematical resource that we employ to compute the inflows every time step, and is therefore unrelated to the diffusion term in Eq. \eqref{eq:sft}, which describes the physical surface diffusion of $B_r$ by the convective flows.

The numerical integration of the diffusion equation for $|B_r|$ or $f$ is done in a number $n$ of time steps that must be large enough for the implicit scheme to produce a reasonably accurate solution and avoid Gibbs oscillations, which lead to instabilities in the simulation. The number $n$ should also be as small as possible to minimize the computational cost. We tested two different approaches to matching this compromise. The straightforward solution is to use time steps of equal size $\delta t' = \Delta t'/n$. By trial and error, we find that $n=50$ satisfies the required conditions. The second approach relies on the fact that the time steps need to be shorter when the gradient of the function to diffuse are steeper. Since the steepest gradients are reduced as the diffusion progresses, taking time steps of increasing size allows us to reduce $n$. Let every time step be longer than the preceding one by a factor $\gamma$. Then:
\begin{equation}
\delta t' = \frac{\Delta t'}{1+\gamma+\gamma^2+\ldots+\gamma^{n-1}}=\frac{1-\gamma}{1-\gamma^n}\Delta t',
\end{equation}
where $\delta t'$ is now the size of the first step.
We found that the combination $\gamma=1.7$ and $n=8$ satisfies the above requirements and leads to results that are consistent with the first approach.

Figure \ref{fig:inflowsExample} shows an example of inflows towards an area of strong magnetic field.

\section{Reference case}\label{sec:refCase}


\subsection{Setup}

We first study the impact of the inflows on the amplitude of the polar fields for our reference values, $a=1.8\cdot10^8\,\mathrm{m^2/G\,s}$ and $\mathrm{FWHM} = 15^\circ$. We ran three sets of simulations: without inflows, with inflows, and with an axisymmetric perturbation of the meridional flow calculated as the azimuthal average of the inflows. The latter is done for comparison with \cite{cameron2012strengths}. While our treatment is not equivalent to the one in the cited study, the calibration factors ensure that the axisymmetric inflows in both studies have similar strengths. To reduce the statistical noise arising from the random positioning of the sources, we ran $20$ realizations for each set of parameters. Each realization spans $55\,\mathrm{years}$.


The initial configuration of the magnetic field is chosen such that the rate of poleward flux transport by the meridional flow and the rate of equatorward transport by turbulent diffusion are approximately equal, $v_m B_r\approx(\eta/R_\odot)\partial B_r/\partial\theta$, corresponding to a situation of activity minimum \citep[see][]{vanBallegooijen1998fchannels}. Combining this with equation \eqref{eq:mFlow} yields:
\begin{equation}
B_r(\lambda) = \begin{cases}
\mathrm{sign}(\lambda)B_0\exp[-a_0(\cos(\pi\lambda/\lambda_0)+1)] &\text{if $|\lambda|< \lambda_0$};\\
\mathrm{sign}(\lambda)B_0 &\text{if $|\lambda|\geq \lambda_0$},
\end{cases}
\label{eq:initialBr}
\end{equation}
where $a_0=v_m R_\odot \lambda_0/\theta\eta$. $B_0$ is chosen in each simulation by requiring that the strength of the polar fields at activity minima remains approximately constant from cycle to cycle.

\subsection{Flux dispersal}\label{sec:fluxDispersal}

Regarding the surface transport of magnetic flux, one of the main questions is how the magnetic flux contained in active regions surrounded by inflows spreads. \cite{derosa2006consequences} found that including the inflows in surface flux transport simulations led to a suppression of the flux dispersal by convective flows, resulting in the formation of magnetic flux clumps incompatible with observations. In \cite{mbelda2015inflows}, we showed that, in the case of an isolated magnetic region, flux cancellation in the first days after emergence causes a decrease in the strength of the inflows, so that turbulent diffusion and the shearing caused by the differential rotation are sufficient to explain the flux dispersal. However, the interaction between neighbouring active regions can result in the formation of polarity-unbalanced magnetic patches. In such cases, flux cancellation does no longer decrease the inflow strength. In our simulations, this leads to the occasional formation of highly concentrated, single-polarity flux clumps which can last for years. This is much longer than the typical decay time of active regions, which ranges from days to weeks \citep[see, e.g.,][]{schrijver_activity}. By contrast, the flux clumping discussed in \cite{derosa2006consequences} possibly arises due to the additional reduction of the diffusivity inside active regions included by these authors in their model. By reducing the diffusivity in active regions, these authors sought to account for the observed reduction of the flux dispersal in the core of active regions \citep{schrijver1990patterns}. In \cite{mbelda2015inflows}, we argued that the inflows alone would have a similar effect.

Figure \ref{fig:tSeries} shows the formation and evolution of one such long-lasting polarity clumps. The first panel shows an active region complex consisting of two large patches of magnetic flux, drawing strong inflows, and a BMR left of it. Some of the negative polarity of the BMR is attracted towards the active complex by the inflows, and cancels part of the complex's positive flux. The positive patch of the BMR, further away from the complex, is less affected and diffuses away rapidly. This causes a flux imbalance in the complex (we stress that magnetic flux is still conserved globally), leading to the formation of the single-polarity feature (top-middle). The persistent inflows further concentrate its flux, until an approximate equilibrium between diffusion and the inflows is reached. As the patch is advected towards higher latitudes, the shear due to differential rotation, the partial cancellation with the polar field, and the artificial decrease of the inflow strength caused by the prefactor $\cos\lambda/\cos 30^\circ$ of our parametrization lead to the dispersal of the single-polarity patch.

\begin{figure}[t]
  \centering
  \resizebox{\hsize}{!}{\includegraphics[width=\textwidth]{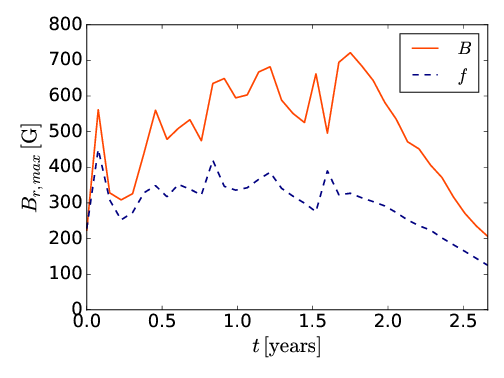}}
  \caption{Maximum field strength of the single-polarity patch under discussion. The orange and dashed, blue lines correspond respectively to the $B-$ and the $f-$ parametrizations of the inflows.}
  \label{fig:clumpFlux}
\end{figure}

The formation and evolution of single-polarity features in simulations using the $f-$parametrization is essentially parallel to the one described for the $B-$parametrization, but the magnetic field is substantially less concentrated. This is demonstrated in Fig. \ref{fig:clumpFlux}, which represents the peak magnetic field of the single-polarity patch as a function of time in both parametrizations. An initial concentration of the magnetic field takes place in the first $\sim2\,\mathrm{months}$ of evolution. This is followed by a dip, possibly caused by the assimilation of positive polarity flux from a decaying BMR, much in the same way the single-polarity feature formed by assimilating mainly one of the polarities of a BMR. A plateau phase follows, in which the average magnetic field of the single-polarity feature, which we may estimate as $\sim B_{r,max}/2$, is $\sim300\,\mathrm{G}$ in the $B-$parametrization case and $\sim150\,\mathrm{G}$ in the $f-$parametrization (although a slightly increasing trend can be seen in the $B-$parametrization case). This phase lasts slightly longer than a year, after which the feature begins to decay owing to the reasons stated above.

The lower concentration of magnetic flux in the long-standing clumps in the $f-$parametrization suggests that a less idealized parametrization of this mechanism may solve the clumping problem completely.

\subsection{Evolution of the axial dipole moment}\label{sec:refCasePol}

\begin{figure}[t]
  \centering
  \resizebox{\hsize}{!}{\includegraphics[width=\textwidth]{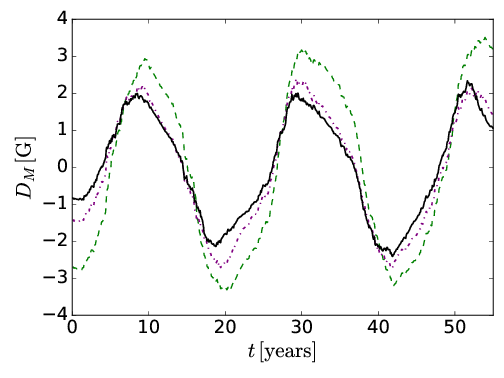}}
  \caption{Evolution of the axial dipole moment in one realization with the $B-$parametrization of the inflows. The (dashed) green, (dotted) purple and black lines correspond to simulations without inflows, with azimuthally averaged inflows and with non-axisymmetric inflows, respectively. Otherwise, the setup of all three simulations is identical.}
  \label{fig:referenceCase}
\end{figure}

To evaluate the impact of the inflows on the reversal and regeneration of the large-scale poloidal field we study the evolution of the axial dipole moment of the surface field, which is defined as:

\begin{equation}
D_M = \sqrt{\frac{3}{4\pi}}\int_0^{2\pi}\int_0^\pi B_r(\phi, \theta) \cos\theta \sin\theta \,\mathrm{d}\theta\mathrm{d}\phi.
\label{eq:admCont}
\end{equation} 

Figure \ref{fig:referenceCase} shows the evolution of the axial dipole moment in three simulations (without inflows, with azimuthally averaged inflows and with full inflows) using the $B-$ parametrization of the inflows. The set of sources is identical in all three realizations. The dipole amplitude is larger in the run without inflows than in the runs with inflows and with azimuthally averaged inflows. In the case with non axisymmetric inflows, $D_M$ is generally lower than in the case with averaged inflows, although values can occasionally be higher due to statistical fluctuations. Averaging the peak values of the dipole moment over 20 realizations yields the following average axial dipole amplitudes:

\begin{align*}
\langle D_M \rangle = 3.27 \pm 0.02 \,\mathrm{G} \;\;&\text{(No inflows);}\\
\langle D_M \rangle = 2.54 \pm 0.02 \,\mathrm{G} \;\;&\text{(Averaged inflows);}\\
\langle D_M \rangle = 2.27 \pm 0.02 \,\mathrm{G} \;\;&\text{(Non-axisymmetric inflows).}
\end{align*}
The amplitude of the axial dipole moment in the case with azimuthally averaged inflows is, on average, a $\sim22\%$ lower than in the case without inflows. In the full-inflows case the average amplitude is $\sim30\%$ lower.

The average dipole peak strength obtained from the simulations using the $f-$parametrization results a $3\%$ lower than in the $B-$case:
\begin{equation*}
\langle D_M \rangle = 2.20 \pm 0.02 \,\mathrm{G} \;\;\text{(Non-axisymmetric inflows, $f$-parametrization).}
\end{equation*}
The slight difference is due to the stronger contribution of fields up to $50\,\mathrm{G}$ in this parametrization (see Sec. \ref{sec:inflowsPar}), which causes a greater restriction of the latitudinal separation of polarities than in the $B-$ case.

The small difference between the average axial dipole moment in the $B-$ and $f-$ cases suggests that the occasional single-polarity clumps do not have a significant impact on the amplitude of the global dipole moment. This is because the single-polarity features occur only very occasionally, so the amount of flux that would have crossed the equator had the feature been allowed to disperse is much smaller than the total flux crossing the equator over the cycle. For this reason, we proceed performing a parameter study of our inflow model using only the $B-$parametrization.

\section{Parameter study}\label{sec:paramStudy}

\subsection{Inflow parameters}

\begin{figure}[t]
  \centering
  \resizebox{\hsize}{!}{\includegraphics[width=\textwidth]{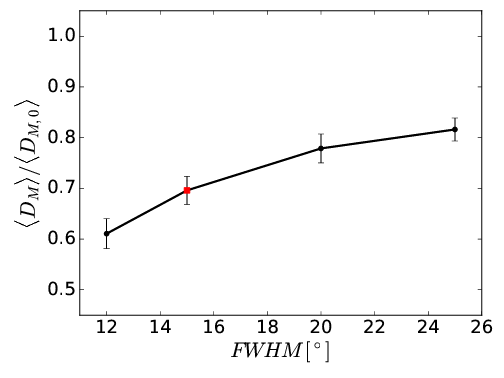}}
  \caption{Average amplitude of the axial dipole moment ($\langle D_M \rangle$) relative to the no-inflows case ($\langle D_{M,0} \rangle$) as a function of the $FWHM$. The red square symbol marks the reference case. The error bars indicate a deviation of $1\sigma$ from the mean.}
  \label{fig:parFwhm}
\end{figure}

\begin{figure}[t]
  \centering
  \resizebox{\hsize}{!}{\includegraphics[width=\textwidth]{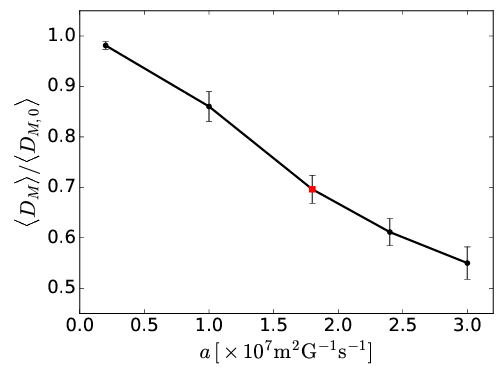}}
  \caption{Average amplitude of the axial dipole moment ($\langle D_M \rangle$) relative to the no-inflows case ($\langle D_{M,0} \rangle$) as a function of the multiplicative parameter $a$ in Eq. \eqref{eq:infVelocParamB}. The red square symbol marks the reference case. The error bars indicate a deviation of $1\sigma$ from the mean.}
  \label{fig:parA}
\end{figure}

To understand the way the two parameters of our model influence the build-up of the axial dipole, we compared the strength of the dipole resulting from simulations with and without inflows. The full width at half maximum of the inflows was varied over the range $12\,^\circ$ to $25\,^\circ$, while keeping $a$ to its reference value of $1.8\cdot10^8\,\mathrm{m^2 G^{-1} s^{-1}}$. Similarly, $a$ was varied from $2\cdot10^7$ to $3\cdot10^8\,\mathrm{m^2G^{-1}s^{-1}}$ with fixed $FWHM=15\,^\circ$. We ran $20$ realizations for each combination of parameters.

Figure \ref{fig:parFwhm} shows the variation of the average dipole peak amplitude with the $FWHM$ of the smoothing Gaussian. For all the values of the $FWHM$, the average amplitude of the axial dipole moment decreases relative to their no-inflows counterpart. This is due to the quenching of the contribution of the BMRs to the global dipole induced by the inflows, which is determined by the magnetic flux of the BMR and the latitudinal separation of the polarities \citep{mbelda2015inflows}. The inflows act to enhance the cancellation of opposite polarity flux and limit the latitudinal separation of the polarities, resulting in a reduction of such contribution. With decreasing $FWHM$, the stronger and more localized inflows further enhance these effects, resulting in a larger reduction of the axial dipole moment. The axial dipole moment ratio varies from $\sim60\%$ to $\sim80\%$ in the considered range of $FWHM$.

Fig. \ref{fig:parA} shows the average peak amplitude of the axial dipole moment as a function of the multiplicative parameter $a$ in Eqs. \eqref{eq:infVelocParamB} and \eqref{eq:infVelocParamF}. For $a=2\cdot10^7\,\mathrm{m^2G^{-1}s^{-1}}$, the average peak amplitude of the axial dipole is only slightly smaller than in the no-inflows case. Larger values of $a$ render stronger inflows, and the amplitude of the dipole consequently decreases. For the strongest inflows considered ($a=3\cdot10^8\,\mathrm{m^2G^{-1}s^{-1}}$), the amplitude of the axial dipole moment is $\sim50\%$ weaker than in the case without inflows.

\subsection{Activity level}

\begin{figure}[t]
  \centering
  \resizebox{\hsize}{!}{\includegraphics[width=\textwidth]{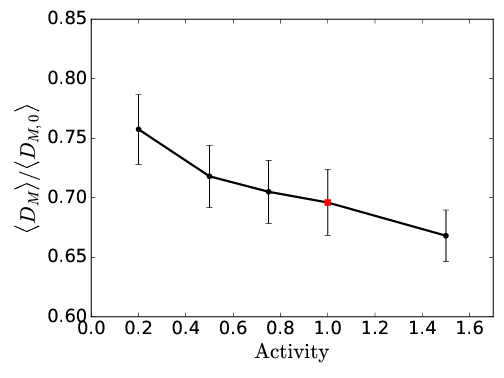}}
  \caption{Average amplitude of the axial dipole moment ($\langle D_M \rangle$) relative to the corresponding no-inflows case ($\langle D_{M,0} \rangle$) as a function of the cycle strength. The symbols indicate the chosen values for the activity level, relative to the activity of the reference case (marked with the red square symbol). The error bars indicate a deviation of $1\sigma$ from the mean.}
  \label{fig:parAct}
\end{figure}

We ran simulations with an activity level, defined as the number of active regions per 11-year cycle, ranging from $0.2$ to $1.5$ times that of the reference case. The dependence of the average peak amplitude of the axial dipole moment (relative to the no-inflows scenario) with activity is shown in Fig. \ref{fig:parAct}. The ratio of dipole moments decreases with the activity level by about a $9\%$ over the whole activity range. This is so because, in strong cycles, due to the larger number of active regions, the collectively driven inflows have a stronger impact on the latitudinal separation of the polarities of individual BMRs. This decrease in the relative amplitude of the axial dipole moment with activity implies, in the Babcock-Leighton framework, that the generation of the poloidal field is more efficient in weak cycles than in strong ones. This constitutes a non-linearity, which may saturate the dynamo and possibly contribute to the modulation of the solar magnetic cycle.

There is a second way inflows can affect the build-up of the axial dipole, namely, enhanced cross-equatorial transport of flux due to inflows driven by low-latitude active regions. This effect is less important in strong cycles than in weak ones, as the former peak earlier than the latter (Waldmeier rule) and, as a consequence, the inflows during the maxima of strong cycles will be further away from the equator. Since all our simulations peak halfway into the cycle, and thus do not include the Waldmeier effect, the influence of the activity level on the build up of the axial dipole may be even stronger than found here.

\section{Conclusion}\label{sec:disConc}
We used a surface flux transport code to study the role of near-surface, converging flows towards active regions on the surface transport of magnetic flux and the build up of an axial dipole at cycle minima. The inflows have been proposed as one possible non-linear mechanism behind the saturation of the global dynamo in the Babcock-Leighton framework \cite{cameron2012strengths}. We stress that other mechanisms, such as alpha-quenching \citep{ruediger1993alpha} or cycle-dependent thermal perturbations of the overshoot region affecting the stability of the flux tubes and, as a consequence, the tilt angle of the emerging flux tubes \citep{isik2015mechanism}, have also been proposed. Here we are concentrating on the inflows, but we do not mean to suggest that in this paper we are excluding other possibilities.

We first studied the evolution of the surface flux in a case with inflows having strength and extension similar to those observed on the Sun. In \cite{mbelda2015inflows}, we found that the strength of the inflows driven by an isolated BMR decays due to the cancellation of opposite-polarity flux over the first $\sim30\,\mathrm{days}$ of evolution. Differential rotation and turbulent diffusion are strong enough to ensure the flux dispersal. However, as seen in section \ref{sec:fluxDispersal}, interaction between neighbouring active regions can occasionally give rise to large single-polarity concentrations. In these cases, a mechanism other than flux cancellation may be required to weaken the inflows and allow for the dispersal of the single-polarity clump. One possibility is that the darkening caused by the formation of pores in areas of strong magnetic field leads to a reduction of the cooling beneath the active region, rendering the inflows weaker. We explored this possibility in our simulations and saw that, although the clumping persists, the magnetic field of these features is substantially lower than in the simulations where the effect of pore-darkening is not considered. This result suggests that this mechanism may be operating in the Sun, although less idealized models of inflows may be necessary to fully account for the clumping problem. In any case, the occasional occurrence of single-polarity clumps in the simulations does not have a significant impact on the amplitude of the global dipole.

We also performed a parameter study in which we varied the strength and extension of the inflows, and the activity of the cycles. In general, inflows decrease the axial dipole moment at the end of the cycle. This is due to the relative decrease in latitudinal separation of the polarities of BMRs caused by the inflows. Stronger (weaker) inflows lead to larger (smaller) reductions of the axial dipole moment. 

Our main finding is that inflows with characteristics similar to those observed can reduce the axial dipole moment at the end of the cycle by a $\sim 30\%$ with respect to the case without inflows in cycles of moderate activity. This ratio varies by a $\sim9\%$ from very weak cycles to very strong cycles, which supports the inflows as a potential non-linear mechanism capable of limiting the field amplification in a Babcock-Leighton dynamo and contributing to the modulation of the solar cycle.

\section*{Aknowledgements}
We want to thank Manfred Schüssler for his valuable suggestions and his thorough revision of this manuscript.

D.M.B. acknowledges postgraduate fellowship of the International Max Planck Research School on Physical Processes in the Solar System  and  Beyond.
 
This work was carried out in the context of Deutsche Forschungsgemeinschaft SFB 963 "Astrophysical Flow Instabilities and Turbulence" (Project A16).

\bibliographystyle{aa}
\bibliography{reviews,papers}

\end{document}